\documentstyle[aps,epsfig]{revtex}
\begin{document}
\wideabs{
\draft \title{Time inhomogeneous Fokker-Planck equation for wave
distributions in the abelian sandpile model} \author{L.
Anton\cite{emailaddress}} \address{Institute for Theoretical Physics,
University of Stellenbosch, Private Bag X1, 7602 Matieland, South
Africa \\ and\\ Institute of Atomic Physics, INFLPR, Lab 22, PO Box
MG-36 R76900, Bucharest, Romania} \date{\today} \maketitle
\begin{abstract}
The time and size distribution of the waves of topplings in the
Abelian sandpile model are expressed as the first arrival at the
origin distribution for a scale invariant, time inhomogeneous
Fokker-Planck equation.  Assuming a linear conjecture for the the time
inhomogeneity exponent as function of loop erased random walk (LERW)
critical exponent, suggested by numerical results, this approach
allows one to estimate the lower critical dimension of the model and the
exact value of the critical exponent for LERW in three dimension. The
avalanche size distribution in two dimensions is found to be the
difference between two closed power laws.
\end{abstract}
 
}

The abelian sandpile model (ASM) was introduced by Bak, Tang, and
Wiesenfeld \cite{Bak87} as a minimal description for natural phenomena
characterized by intermittent time evolution through events called
avalanches which have scale invariant properties. The model is defined
on a hypercubic lattice whose sites can accommodate a variable,
positive number of grains. With a uniform distribution a site is
chosen and its number of grains is increased by one. If its total
number of grains exceeds a given critical threshold $z_{max}$, the
nearest neighbor sites increase their number of grains by one and the
initial site loses the corresponding number of grains. Then the newly
updated sites are checked for stability until there are no more
unstable states. This event is an avalanche and the analytical
properties of its distribution is still an open question
\cite{Ktitarev00,Menech98,Menech00,Drossel99,Tebaldi99}. Analytical
approaches rely on the algebraic properties of the toppling rules and
the decomposition of avalanches into simpler events called waves which
are related to spanning trees on a
lattice\cite{Ktitarev00,Dhar90,Ivashkevich94}. A wave of topplings is
simply obtained by restraining the initial site, from where an
avalanche was initiated, to topple again only after all the other
unstable sites have relaxed.  Recently it was shown that the wave
distribution satisfies the finite size scaling ansatz, with critical
exponents deduced from the equivalence between waves and spanning trees
\cite{Ktitarev00}.

In this Letter we present numerical evidence that the critical
exponents of the wave size and time distribution, which were deduced
from geometrical considerations in
\cite{Ktitarev00,Dhar90,Ivashkevich94}, can be related to the
parameters of a scale invariant Fokker-Planck equation (FPE) in any
dimension. In this way we make a connection between the geometrical
properties of the waves and an evolution equation. As further
confirmation of the validity of a FPE description for the abelian
sandpile model, we show that the scaling behavior of the average
number of unstable sites as function of time is predicted by the
FPE. Using the expressions for the critical exponents deduced in
\cite{Ktitarev00} together with those inferred form the FPE approach
we are able to find the lower critical dimension of the abelian
sandpile model and, as an extra benefit, the exact value for the loop
erased random walk (LERW) critical exponent $\nu$ in three
dimensions. In two dimensions, the case in which the distribution
of the last wave is known \cite{Ivashkevich94}, we can compute the
asymptotic behavior of the avalanche distribution with the result that
it has the form of a difference between two close power laws. This has
been previously proposed as an explanation for the failure of the
finite size scaling approach \cite{Menech98,Dhar99}.

The abelian sandpile model is, from its definition, a Markovian
process whose states are specified by the lattice configuration. Once
the initial point has been chosen randomly the dynamics of relaxation is
 deterministic, with the  evolution determined  by the
initial configuration and the toppling rule.
 
We consider a coarse grain description of the sandpile
evolution. Instead of a complete description, involving the number of
grains at each site of the lattice, we use as variable the total
number of unstable sites which exist at a given time, after an
avalanche has started, irrespective of the configuration in which the
sandpile is.  This description is stochastic since the transition
between two states with a given number of unstable sites depends on
the configuration of the sandpile which is now taken randomly.  Let us
consider for this process the evolution equation describing the
transition between states with different numbers of unstable sites: $
P(t+1, n)=\sum_{n'} W(t;n,n')P(t,n')$; where $ W(n,n';t) $ is obtained
by averaging over all the transitions between the configurations with
the same number of unstable sites $n,n'$ at a given time $t$. The
transition probability $W(t;n,n')$ may depend on time since the
configuration of the unstable sites also depends on time; i.e. at the
first step of a wave the unstable sites are some of the nearest
neighbors of the initial site, eventually moving away.  In this
formulation the wave is equivalent to a particle performing a discrete
random walk on the positive semi-axis with the transition
probabilities depending on the particle position.  An wave event
is a first arrival at the origin problem, for the wave stops when all
the sites, except the initial one, are stable.  This random walk is
close to a diffusion process since the number of unstable sites varies
with bounded steps, $x(t+1)<2Dx(t)$, where $x(t)$ is the number of
unstable sites and $D$ is the lattice dimension.  From this analogy we
expect that the distribution of first arrival at the origin, that is
the wave duration, has a power law distribution as it has for the
first return distribution of the simple diffusion process.

\begin{figure}
\vspace*{-15mm}
\centering
\epsfig{file=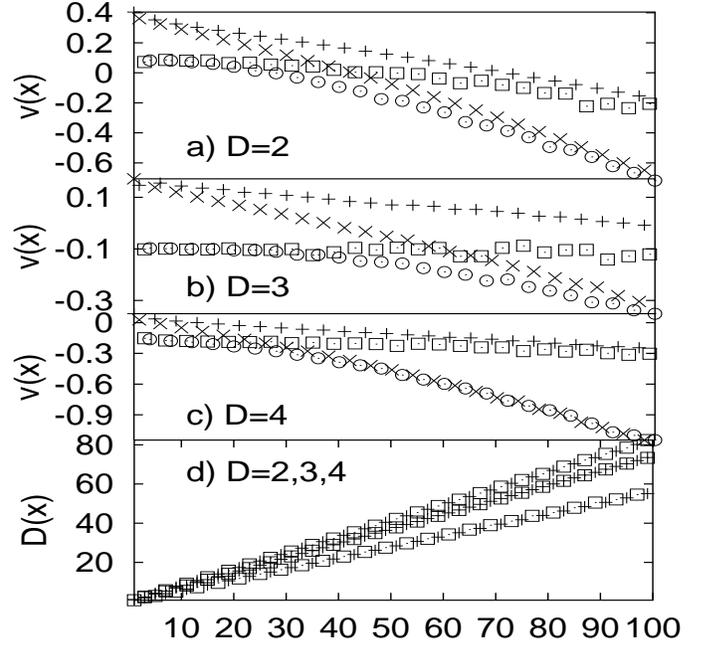,%
height=100mm, width=75mm,bbllx=100pt,bblly=100pt,bburx=495pt,bbury=742pt}
\vspace*{5mm}
\caption{The local drift coefficient for bulk and boundary wave in
a)$D=2$; $L=1024\,(+)$, $L=512\,(\times)$ for bulk waves and
$L=1024\,(\Box)$, $L=512(\circ)$ for boundary waves; b) $D=3$;
$L=128\,(+)$, $L=64(\times)$ for bulk waves and $L=128\,(\Box)$,
$L=64(\circ)$ for boundary waves; c) $D=4$; $L=32\,(+)$, 
$L=16\,(\times)$ for bulk waves and $L=32\,(\Box)$, $L=16(\circ)$ for
boundary waves; d)the local second order moment for bulk waves $(+)$
($D=2(L=1024)$, $D=3(L=128)$, $D=4(L=32)$) and for boundary waves
$\Box$ ($D=2(L=512)$, $D=3(L=64)$, $D=4(L=16)$). 
}
\label{fig1}
\end{figure}

If we take the lattice size to infinity and the time unit to zero in
an appropriate way the discrete Markov chain can be cast into a
diffusion equation via the Kramer-Moyal expansion \cite{Risken96}.
Here we shall not propose an explicit way to construct the FPE for the
ASM, instead we use the fact that in the stationary state the sandpile
is at criticality and we shall investigate the general FPE which
yields critical behavior and has diffusion and drift coefficients
behaving similarly to the sandpile model.  Generically the FPE has the
form $ \partial_t p =-\partial_x [v(x,t)p]
+(1/2)\partial^{2}_{xx}[D_2(x,t)p] $, where $p(x,t)$ is the
probability density of the number of unstable sites, $x$ is the number
of unstable sites, $t$ is the time since the wave has started. The
drift coefficient $v(x,t)$ and the diffusion coefficient $D_2(x,t) $
are obtained by taking the continuum limit of the local first order
moment $\sum_j(x_i-x_j)W(i,j;t)$ and of the local second order moment
$\sum_j(x_i-x_j)^2W(i,j;t)$ \cite{Risken96}.  Numerically we have
found that the discrete diffusion coefficient $D_2(x,t)$ depends
linearly on the number of unstable sites $x$. The slope depends on the
dimension of the lattice but does not depend on the lattice size and
the geometric condition (bulk or boundary wave)( see
Fig. \ref{fig1}).  Also we have found that the discrete drift
coefficient $v(x)$ tends to a constant as the size of lattice grows,
Fig.\ (\ref{fig1}). The finite size effects affect the drift
coefficient for the bulk waves at any value of the number of unstable
sites, since the transition to states with larger number of unstable
sites is smaller when the wave takes place near the boundary, while
the statistics collects all the waves.

The simplest Fokker-Planck equation satisfying the scale invariance
assumption and the numerical behavior of the discrete coefficients
$D_2(x,t)$ and $v(x,t)$ is
\begin{equation}\label{FPinit}
\partial_t p(x,t)=
-\partial_x[v t^{-\alpha} p(x,t)]
+\partial^{2}_{xx}[D_2 x t^{-\alpha} p(x,t)]
\end{equation}   
where $v,D_2,\alpha$ are constants. The initial condition for the
 above equation is $p(x,t=t_0)=\delta(x-x_0)$. Since we are interested
 in the time and size distribution for waves, which are first arrival
 events, we set an absorbing boundary condition at the origin
 $p(x=0,t)=0$, for the wave stops when the number of unstable sites,
 except for the initial one, is zero.  The above differential equation
 is invariant under a scale transformation $x\rightarrow bx,\quad
 t\rightarrow b^{1/(1-\alpha)}t$. We observe that we can eliminate the
 parameter $D_2$ by variable change $x\rightarrow D_2x$ and we change
 to $v\rightarrow v/D_2$.

Using a standard approach \cite{Feller71b} one can find the asymptotic
behavior of the first arrival at the origin for Eq. (\ref{FPinit}):
 $P_t(t)\approx t^{-\tau_t},\quad t\gg 1$ with
\begin{eqnarray}
\tau_t&=&1+(1-\alpha)|1-v|\label{eqtaut}.
\end{eqnarray}

The second critical exponent we are interested in is the size
distribution of waves. The size of the wave is the sum of the
number of  unstable sites until the wave stops; in the continuous
formulation we have $ s(t)=\int_{t_0}^{t}dt' x(t')+x(t_0) $.  

 We make the observation that the variable $s$ is a monotonic function
 of time as $\dot{s}=x(t) > 0$.  The relation between the two
 variables can be found using the average relation $ \langle \dot
 s(t)\rangle=\langle x(t)\rangle$.  Multiplying Eq. (\ref{FPinit}) by
 $x$ and integrating over $x$ we obtain that $ \langle x(t)\rangle
 \approx t^{1-\alpha},\quad t \gg 1 $, where the average was
 normalized to the probability of surviving $ \int dx p(x,t)$ until
 the moment $t$.  Now we can use the variable $s$ in the time
 distribution of waves.  We have $t^{-\tau_t}dt\approx
 s^{\frac{1}{2-\alpha}}ds^{\frac{1}{2-\alpha}}=s^{-\tau_a}ds$ for
 large $t$ and $s$, where
\begin{equation}
\tau_a=1+\frac{(1-\alpha)}{2-\alpha}|1-v|\label{eqtaua}.
\end{equation}
This result can be checked easily by Monte Carlo simulation of a
random walk constructed to be the discrete version of
Eq. (\ref{FPinit}) or equivalent ones.

In Table \ref {table} we show the values of the critical exponents
$\tau_a,\tau_t$ taken from \cite{Ktitarev00}, using
Eqs.~(\ref{eqtaut},\ref{eqtaua}) we have computed the values of the
parameters $v$ and $\alpha$. We observe that $\alpha$ has the same
value for bulk and boundary waves and the time inhomogeneity
disappears at the critical dimension, $\alpha=0$ for $D=4$. Thus the
exponent $\alpha$ can be interpreted as a measure of the correlation
among the unstable sites bellow critical dimension. 

Also, for $D=2$ we note that $\alpha$ has the same numerical value as
the critical exponent characterizing the decay of the autocorrelation
function of waves found in \cite{Menech00}.  One more hint that the
exponent $\alpha$ is related to the correlation on the lattice is that
it depends linearly on the erased random critical exponent $\nu$ which
has the values $\nu=4/5\,( D=2),\, \nu=0.616\, (D=3),\, \nu=1/2\,
(D\ge 4)$ \cite{Ktitarev00,Bradley95}. Indeed one can see easily that
the relation
\begin{equation}\label{eqaofnu}
\alpha=\frac{4}{3}\Bigl(\nu-\frac{1}{2}\Bigr)
\end{equation}
holds exactly in $D=2$ and $D=4$ and has an error of $0.003$ for
$D=3$, case in which the critical exponent $\nu$ is only known from
numerical simulation \cite{Bradley95}. 

\begin{figure}
\centering
\epsfig{file=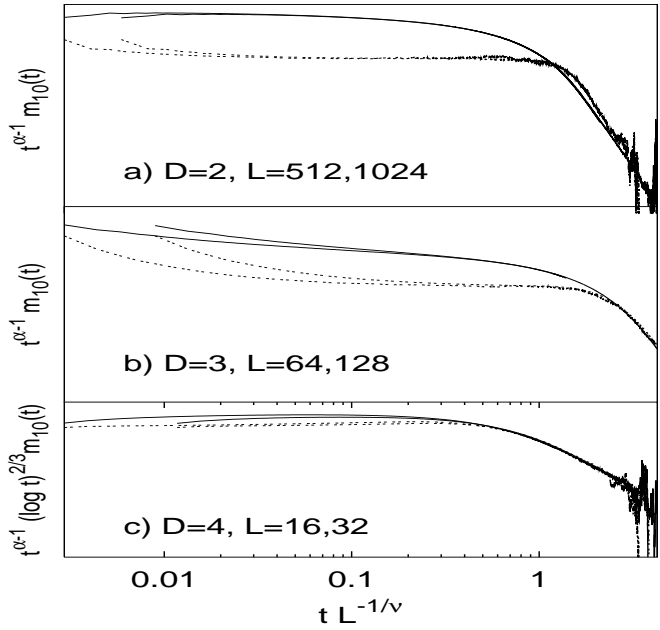,%
height=80mm, width=75mm,bbllx=100pt,bblly=100pt,bburx=495pt,bbury=742pt}
\vspace*{5mm}
\caption{Finite size scaling for the moment ratio $m_{10}(t)=m_1(t)/m_0(t)$
in: $D=2$(a), $D=3$ (b), $D=4$ (c) with continuous line for bulk
waves and with dashed line for boundary waves. There is a good fit of the
data with the exponents taken from Table (\protect\ref{table}). The
logarithmic correction to scaling at $D=4$ has the same exponent as
in Ref.\ \protect\cite{Ktitarev00}.}
\label{fig2}
\end{figure}

In the following we shall use Eq. (\ref{eqaofnu}), which holds for
both bulk and boundary waves, as a conjecture for further exploration
of the Table I. The fact that $\alpha$ has the same value for both
bulk and boundary waves can be checked numerically using the first two
moments $m_n(t)=\int x^np(x,t)$ of the solution of
Eq. (\ref{FPinit}). Integrating Eq.\ (\ref{FPinit}) over $x$ and using
the absorbing boundary conditions at the origin, we obtain
$m_1(t)/m_0(t)\approx t^{1-\alpha}$ which is independent of $v$.
Numerical estimation of the above ratio in $D=2,3,4$, presented in
Fig. (\ref{fig2}), shows an excellent agreement with the predicted
values (the error is less than $0.01$).

In \cite{Ktitarev00} it was shown that for waves the critical
exponents for size and time distribution are given by
\begin{equation}\label{tauktit}
\begin{array}{l}
\tau_{a}=2-\frac{1+\sigma}{d_f}\\
\tau_{t}=1+(d_f-(1+\sigma))\nu
\end{array}
\end{equation} 
where $d_f$ is the fractal dimension of the wave, which has the values of
the Euclidean dimension for $D=2,3,4$ and value $4$ for higher
dimension; $\sigma=1,0$ for bulk and boundary
wave respectively. 

Using the above proposed conjecture, Eq.~(\ref{eqaofnu}), and
Eqs.~(\ref{eqtaut},\ref{eqtaua},\ref{tauktit}), from
$(2-\alpha)(\tau_a-1) =\tau_t-1$ we obtain
\begin{equation}\label{eq:d_f}
d_f=\frac{8}{3}\frac{1}{\nu}-\frac{4}{3}.
\end{equation}
This relation shows that that the minimum critical dimension of the
abelian sandpile model is $4/3$ since the maximum value of $\nu$ is
$1$. This result is in agreement with the observation that in $D=1$
the scaling in the abelian sandpile model breaks down \cite{Dhar99}.

An additional benefit of the above relation is that it gives the value
of the critical exponent for the loop erased random walk in three
dimension. Indeed if we put $d_f=3$, and keep in mind that $d_f=D$
for $D<4$, we get $\nu_{D=3}=8/13=0.615384...$ in perfect agreement
with the numerical value found in \cite{Bradley95}. In fact, assuming
the conjecture Eq. (\ref{eqaofnu}) and identifying $D=d_f$,
Eq. (\ref{eq:d_f}) yields a relation between the LERW critical exponent
and space dimensionality.

Ktitarev {\it et al} \cite{Ktitarev00} argue that all critical
exponents of the abelian sandpile are determined by the critical
exponent $\nu$. In order to complete this program we need the relation
between the drift coefficient $v$ and $\nu$ and a relation between the
the drift coefficients $v$ for the bulk and boundary waves. Using the
Eq. (\ref{tauktit}) for $\tau_t$ together with
Eqs. (\ref{eqtaut},\ref{eqtaua}) we obtain 
\begin{equation}\label{vofv}
\frac{|1-v_{\text{bulk}}|}
{|1-v_{\text{boundary}}|}
=\frac{d_f-2}{d_f-1}.
\end{equation} 

We need one more relation to connect the coefficients $\alpha$ and $v$
(bulk or boundary) with $\nu$. This can be obtained by using for
example Eqs. (\ref{eqtaua},\ref{eqaofnu},\ref{tauktit}) from which we
extract
\begin{equation}\label{vofnu}
v_{\text{bulk}}=\left\{
\begin{array}{ll}
\frac{6\nu-3}{-4\nu+5}&\quad \mbox{if}\quad \nu \le 4/5\\
\frac{-14\nu+13}{-4\nu+5}&\quad \mbox{if}\quad \nu > 4/5\\
\end{array}.
\right .
\end{equation} 

We make the observation that the
Eqs.~(\ref{eqaofnu},\ref{vofv},\ref{vofnu}) hold exactly in $D=3$ if
we use the deduced value $\nu=8/13$, see Table I.  At this point, we
can conclude that we have found a self-consistent description of the
critical properties for the time and size distribution of waves in
ASM.  The exponent $\alpha$ captures the lattice correlation and the
drift $v$ controls the boundary condition of the wave, both of them
being a function of the erased loop random walk critical exponent
$\nu$ through the Eqs.~(\ref{eqaofnu},\ref{vofv},\ref{vofnu}).

Now we are in the position to compute the asymptotic behavior for the
avalanches in $D=2$ using the FPE. In this description an avalanche is
the sum of a random number of waves. The waves are statistically
independent being recurrent events \cite{Feller71a}. This assumption
might appear to be in contradiction with analysis of \cite{Menech00}
but in fact in this approach the correlation is included already in
the time inhomogeneity. When a wave of size $s$ touches the origin it
has the the probability $p_d(s)$ to die, thus also concluding the
avalanche, or a new wave can start with the probability
$(1-p_d(s))$. We choose the probability for the wave to die as
$p_d(s)=s^{-3/8}\ln s$, so as to recover asymptotically the
probability for the last wave: $p_w(s)p_d(s)= p_{lw}(s)\sim
s^{-\frac{11}{8}}$ \cite{Dhar94,Ivashkevich94}. The avalanche
distribution can then be written as
$
p_{a}(s)=p_w(s)p_d(s)
+\int_{0}^{s}ds' p_{w}(s')(1-p_d(s'))
p_{w}(s-s')p_d(s-s')+\dots
$.
We can sum the previous series after a Laplace transform in $s$ and we
have
\begin{equation}
p_a(\lambda)=\frac{p_{lw}(\lambda)}{1-((1-p_d)p_w)_\lambda}.
\end{equation}
Applying again the Tauberian theorem \cite{Feller71b} we find that
asymptotically the avalanche size distribution behaves like
\begin{equation}
p_a(s)\approx C_1(s\ln s)^{-1}+C_2 s^{-\frac{11}{8}}.
\end{equation}
This kind of behavior has been proposed previously in the literature
\cite{Dhar99}. The fact that $1< 11/8 <2$ makes it difficult to obtain
the 'pure' dominant behavior. From a numerical fit we obtain that
$C_2/C_1\approx -0.25$, therefore $C_1(s\ln s)^{-1} \gg C_2
s^{-\frac{11}{8}}$ for $s \ge 10^6$. Thus, the FPE approach predicts
that the avalanche distribution in the bulk must have the same
asymptotic behavior as the waves for very large values of $s$,
provided that the statistics excludes the avalanches which are
affected by the boundary.

In conclusion, we have used numerical hints to propose a FPE for the
time and size distributions of the waves in the ASM.  In this approach
a wave is a first return event and the asymptotic properties of its
distributions (time and size) are described by the first return
probabilities of a time inhomogeneous FPE; the time inhomogeneity
appears below the critical dimension $D=4$. Furthermore, this
approach yields an analytical expression for the asymptotic behavior
of the avalanche distribution in $D=2$ which goes beyond the finite size
scaling hypothesis and in agreement with recent results
\cite{Menech98,Dhar99}.

Using the relation for the critical exponents $\tau_a$, $\tau_t$
deduced in \cite{Ktitarev00} together with the the relations found
through FPE approach and the conjecture (\ref{eqaofnu}) we the propose
explicit dependence of the critical exponents $\tau_a$, $\tau_t$ of
the critical exponent of LERW $\nu$ (via $\alpha$ and $v$). A bonus of
this approach is that it yields the value of the lower critical
dimension, $4/3$, for the ASM and the exact value of $\nu=8/13$ in
$D=3$ for LERW.

The author thanks H.~B.~Geyer, F.~Scholtz, L.~Boonaazier and A.~van
~Biljon for useful discussions and H.~B.~Geyer for a critical reading
of the manuscript.

\begin{table}
\caption{The values for the critical exponent $\tau_t,\,\,\tau_a$ for
the time and size distribution of waves taken from
\protect\cite{Ktitarev00} together with the values of the exponent
$\alpha$ and the drift coefficient $v$ computed from Eqs.
(\protect\ref{eqtaut},\protect\ref{eqtaua}). The last line shows the
values of the erased loop random walk critical exponent $\nu$. For
$D=3$  we show in parenthesis the values computed with the exact value
 $\nu=8/13$.}
\label{table}
\begin{tabular}{llll}
D & 2 & 3 & 4\\
\tableline
bulk& $\tau_a=1$& $\tau_a=4/3$ & $\tau_a=3/2$\\
    & $\tau_t=1$& $\tau_t=1.616\;(19/13)$ & $\tau_t=2$\\
    & $\alpha=2/5$& $\alpha=0.152\;(2/13)$ & $\alpha=0$\\
    & $v=1$      & $v=0.274\;(3/11)$ & $v= 0$\\
boundary& $\tau_a=3/2$& $\tau_a=5/3$ & $\tau_a=7/4$ \\
    & $\tau_t=9/5$& $\tau_t=2.232\;(29/13)$ & $\tau_t=5/2$ \\
    & $\alpha=2/5$& $\alpha=0.152\;(2/13)$ & $\alpha=0$ \\
    & $v=-1/3$      & $v=-0.453\;(-5/11)$ & $v=-1/2$\\
$\nu$ & $4/5$ & $0.616\;(8/13)$ & $1/2$
\end{tabular}
\end{table}

\end{document}